# Optical information capacity of silicon

Dimitris Dimitropoulos, Bahram Jalali[1,2,3]

[1] Department of Electrical Engineering, University of California, Los Angeles, 420 Westwood Plaza, Los Angeles, USA 90095

[2] Department of Bioengineering, University of California, Los Angeles, 420 Westwood Plaza, Los Angeles, USA 90095

[3] California NanoSystems Institute (CNSI), 570 Westwood Plaza, Los Angeles, CA 90095

**Abstract** Modern computing and data storage systems increasingly rely on parallel architectures where processing and storage load is distributed within a cluster of nodes[1,2]. The necessity for high-bandwidth data links has made optical communication a critical constituent of modern information systems and silicon the leading platform for creating the necessary optical components[3,4]. While silicon is arguably the most extensively studied material in history, one of its most important attributes, an analysis of its capacity to carry optical information, has not been reported. The calculation of the information capacity of silicon is complicated by nonlinear losses, phenomena that emerge in optical nanowires as a result of the concentration of optical power in a small geometry. Nonlinear losses are absent in silica glass optical fiber and other common communication channels. While nonlinear loss in silicon is well known[5,6], noise and fluctuations that arise from it have never been considered. Here we report sources of fluctuations that arise from two-photon absorption and free-carrier plasma effects and use these results to investigate the theoretical limit of the information capacity of silicon. Our results show that noise and fluctuations due to nonlinear absorption become significant and limit the information capacity well before nonlinear loss itself becomes dominant. We present closed-form analytical expressions that quantify the capacity and provide an intuitive understanding of the underlying interactions. Our results provide the capacity limit and its origin, and suggest solutions for extending it via coding and coherent signaling. The amount of information that can be transmitted by light through silicon is the key element in future information systems. Results presented here are not only applicable to silicon but also to other semiconductor optical channels.

The bedrock of the electronics industry, silicon, is now recognized as a photonic platform for creating components that provide data communication for massive arrays of servers and routers in data centers[1-4], because it solves the interconnect bottleneck at the board and eventually at chip scale[5]. Motivations are economic as well as performance. Silicon photonics makes use of the same high-volume manufacturing infrastructure that has led to the explosive growth of consumer electronics and the internet. Moreover, optical interconnects offer lower delays, higher bandwidth, and immunity to electromagnetic interference[7].

The amount of information that can be transmitted through silicon is a fundamental question that has so far not been considered. Shannon's classical theory of information theory[8] shows that the capacity of a channel increases with signal power because, in the presence of ambiguity caused by noise, the number of distinguishable signal levels increases. But as we will show, the concentration of power in limited cross sections activates new types of noises that place a fundamental limit on the optical information capacity of silicon. This predicament is emphasized by the economics of silicon manufacturing, requiring efficient use of chip area[9] that drives down the optical wire or waveguide cross section to near the diffraction limit of light (0.25-0.5 $\mu m^2$).

The information capacity of a communication channel is the highest rate at which information can be transmitted reliably through the channel. For a simple channel where transmission is linear and the only source of noise is an additive Gaussian source with power $N$, the capacity, measured in bits is $C = \log_2(1 + S/N)$ where $S$ is the signal power[10]. As a reference, for a linear channel with signal to noise ratio of 30 dB, the capacity is approximately 10 bits. Currently, electro-optic modulators and electronics limit the bandwidth to about 30 GHz (for 40 Gbps data), suggesting a capacity of 300 Gbps. This can be increased using multiple wavelength channels, each operating at the channel capacity. Multiple wavelengths within a single waveguide also improve silicon real estate usage, an important consideration in the economics of silicon manufacturing[9,11]. As long as the channel is linear, the capacity can be improved by increasing the signal power and the number of wavelength channels.

Optical media exhibit nonlinearities at high optical powers and this places an upper limit on the capacity that can be predicted by the linear channel model[12-19]. In the case of optical fiber, it has been shown that cross phase modulation, arising from the inherent 3rd order nonlinearity present in the glass medium (Kerr effect) places a limit on the capacity of optical fibers[12]. This important result was in contrast to the classic linear channel where capacity monotonically increases as a function of input power.

Silicon photonic devices work with infrared light (wavelength in the 1,500 nm range) where the photon energy is insufficient to be absorbed. However at high intensities, two photons will pool their energies together to effect two-photon absorption (TPA)[5,6]. The rate, $G$, of this nonlinear absorption process depends on the light intensity as $G = \beta I^2/2hf$, where $I$ is the intensity and $\beta \sim 0.5$ cm/GW is the two-photon absorption coefficient. The corresponding loss coefficient is $a_{TPA} = \beta I$, measured in inverse centimeters ($cm^{-1}$). TPA also initiates a plasma effect: electrons and holes generated by TPA cause free-carrier absorption (FCA), the magnitude of which is proportional to the density of carriers, $n_C = G\tau$, where $\tau$ is the carrier lifetime. This plasma effect in the form of the so-called FCA loss is characterized by a loss coefficient $a_{FCA} = \sigma n_C$, where $\sigma \sim 10^{-17}$ $cm^2$ is the FCA cross section[6,20]. But TPA and FCA cause fluctuations – important phenomena that have not been considered so far. These render the calculation of the information capacity of silicon significantly more complicated than that of optical fibers.

The refractive index $n$, depends on the intensity in the form of $n = n_o + n_2 I$ [21]. Here, $n_o$ is the linear index and $n_2$ is a constant characterizing the strength of the nonlinearity. In multi-channel or wavelength division multiplexing (WDM) communication, the phase in a given channel is influenced by intensities in other channels by a process called cross-phase modulation (CPM), and therefore coherent communication is affected (four-wave mixing is of less concern as discussed in the Methods section).

In long-haul communication, optical amplifiers are used to compensate for the loss of long fiber spans. The noise floor is then determined by the amplified spontaneous emission (ASE) of the amplifier[21]. Due to practical and economic reasons, optical amplifiers are not standard elements in silicon photonics. The principal noise sources in silicon photonics, aside from noise due to nonlinear loss, are then thermal noise and shot noise.

As we will show, TPA and FCA nonlinearities have profound implications for information transmission in silicon waveguides, even at low intensity levels where the optical transmission is barely affected by the nonlinear effects. TPA is an instantaneous process because the nonlinear loss does not depend on past transmitted symbols. FCA, on the other hand, is not, since the carriers, generated due to TPA, do not recombine immediately but are present in the waveguide for the duration of the carrier lifetime. Therefore, TPA only adds noise in WDM intensity modulation schemes, where each transmitted channel is a source of instantaneous (fluctuating) loss for the others. FCA, on the other hand, adds noise in WDM, as well as in single-channel systems. For WDM, our results show that noise due to TPA is responsible for the value of the peak capacity that can be achieved (as well as the mean input optical power at which the maximum is achieved). FCA noise determines the rate of capacity decrease with power beyond the maximum.

First let us consider only the presence of TPA in a WDM scheme where $N_c$ channels are being transmitted, each with power $P_j(z)$ along the dimension $z$ of a waveguide of total length $L_{wg}$. The total power causes a modulation of the transmission factor of every channel through the waveguide. In a WDM system with a large number of separately coded channels, $N_c \gg 1$, each channel sees random multiplicative noise due to the other channels (see Methods section for details).

FCA in single-channel intensity-modulated transmission creates a similar problem. Any carrier density generated due to TPA in the waveguide has a characteristic decay lifetime, $\tau$. Therefore, the carrier density present at any given time depends on the signal intensity during a past time interval equal to the carrier lifetime. When the pulse (symbol) duration $T_s$ is much shorter than the lifetime, i.e., $\tau/T_s \gg 1$, as is typically the case, the number of past symbols that contribute to the FCA loss of the current pulse is $\approx \tau/T_s$ (see Methods section for the statistics of the loss).

We first consider transmission that employs intensity modulation of the optical signal. The input random variable $X$ ($X \geq 0$) is the optical intensity at the input of the waveguide and the output $Y = LX + N$ is the measured photocurrent. Here the random variable $N$ describes the thermal and shot-noise current at the photodetector, and $L$ ($L \geq 0$) is the channel transfer function that includes TPA and FCA nonlinear losses. It can be approximated with a log-normal random variable $L = e^{-l+sZ}$, where $l$ is the total loss coefficient, $Z$ is a zero mean normal variable with unity standard deviation, and $s$ is the standard deviation of noise caused by TPA and/or FCA. Please see the Methods section for a description of the channel parameters for: (a) a WDM system where TPA noise sets the limit and for (b) single-channel transmission where only FCA noise is dominant.

To evaluate the maximum information that can be transmitted by a channel, traditionally we take the view that the input $X$ and output $Y$ are random variables. The mutual information $I(Y;X)$, is a measure of the dependence between the input and the output, and is given by $H(X;Y) = H(Y) - H(Y|X)$, where $H(Y) = -\langle \log p(Y) \rangle_Y$ and $H(Y|X) = -\langle \log p(Y|X) \rangle_{Y,X}$ with $H$ denoting the information entropy. The maximum of $I(Y;X)$, over the possible input probability distributions, is the bits per channel use that can be reliably transmitted through a channel[10].

The following lower limit on the capacity is obtained from a lower bound on $H(Y)$ through application of the entropy power inequality[10], and from an upper bound on $H(Y|X)$ that is obtained by noting that the random variable $Y|X$ has a lower entropy than a Gaussian variable of the same variance:

$$I_{LB} = \frac{1}{2}\ln\left(e^{2H(N)} + e^{2\langle \ln X \rangle + 2\langle \ln L \rangle} + e^{2H(X) + 2\langle \ln L \rangle}\right)$$

$$-\frac{1}{2}\ln(2\pi e) - \frac{1}{2}\langle \ln(\langle N^2 \rangle + \langle \Delta L^2 \rangle X^2) \rangle \quad (1)$$

Using the method of Lagrange multipliers, this distribution can be shown to have the form:

$$p(X) = \frac{CX^{a-1}e^{-\lambda X}}{(\langle \Delta L^2 \rangle X^2 + \langle N^2 \rangle)^{b/2}}, \quad C, a, b, \lambda > 0 \quad (2)$$

where the parameters $C, a, b, \lambda$ are dependent on the mean of the input $\langle X \rangle$. Optimizing the bound numerically shows that nearly maximal results are obtained with an exponential distribution ($a = 1$) resulting in the following lower bound for intensity-modulated WDM transmission:

$$C_{LB}(\text{bits/s} \cdot \text{Hz}) \cong \frac{1}{2}\log_2\left[\frac{1 + \frac{e^{1-2l}}{2\pi}\left(\frac{P}{P_N}\right)^2\left(1 + 2\pi e^{-1-2\gamma}\left(\frac{P}{P_{TPA}}\right)^2\right)}{1 + 2e^{-2l}\left(\frac{P}{P_N}\right)^2\left(\frac{P}{P_{TPA}}\right)^2}\right] \quad (3)$$

Here $P = \langle X \rangle$ is the average optical power per channel, $P_N^2 = \langle N^2 \rangle$ is the noise power as measured in electrical domain, $l$ is the total mean loss coefficient along the waveguide and $P_{TPA} = \frac{A_{eff}}{N_C^{1/2}\beta_{TPA}L_{eff}}$ is a characteristic power above which the transmission begins to be impaired by TPA and $\gamma = 0.577$. The additive noise power contains both thermal and shot noise terms, $\langle N^2 \rangle = (\hbar\omega/q)^2(kT/R +$ $2q(q/\hbar\omega)\langle X\rangle)W$ . The peak of the capacity is reached at a power level $P_{peak} = (e^l P_{TPA} P_N)^{1/2}/2^{1/4}$. The optical power at which the capacity is reached is typically high enough for shot noise to dominate thermal noise. In that case, $P_{peak} \cong (e^{2l} P_{shot} P_{TPA}^2/2)^{1/3}$ , where $P_{shot} = 2\hbar\omega W$ and the peak capacity is approximately $C_{peak} \cong \frac{1}{3}\log_2\left(\frac{e^{-2l} P_{TPA}}{P_{shot}}\right) + \frac{1}{2}\log_2(e/(2^{1/3} 4\pi))$ .

A similarly simple result is detailed in the Methods section for a single channel limited by FCA.

So far we have discussed intensity modulation because it is the domain where silicon photonics operates. The situation is simpler in coherent communication because the lack of intensity modulation eliminates fluctuations due to TPA and FCA. This limit is then due to the Kerr effect in silicon and is discussed in the latter part of this letter.

The TPA induced crosstalk penalty on the capacity is illustrated in Fig. 1 for a set of typical parameters. The linear regime of increasing capacity is interrupted when the TPA-induced crosstalk sets in at slightly less than 35 dB of optical SNR (70 dB electrical). After that point, the capacity decreases with the same rate that it was previously increasing. The reason for this is not the nonlinear loss, but rather the induced crosstalk. Capacity is a measure of dynamic range, and in the nonlinear regime an increase in the input power translates in an even higher increase of the crosstalk level. The maximum capacity attained is about 9.4 bit/s/Hz. For comparison we also show : (i) our channel lower bound when only additive thermal noise is present at the receiver $L = 1$, (with signal input that follows the exponential distribution), where the capacity keeps increasing with increasing power $C_{awgn,non-coh} = \frac{1}{2}\log_2\left(1 + \frac{e}{2\pi}\frac{P^2}{P_N^2}\right)$, (ii) ½ of the coherent capacity bound of Mitra-Stark (iii), ½ of the AWGN band-limited channel capacity.

The SNR and power dependence of these results merits some discussion. In the coherent case shot-noise limited detection is assumed (with unity photon/electron conversion efficiency), with an additive noise electrical SNR, $SNR_e = P/(\hbar\omega W)$, at the photodetector. This SNR can always be achieved in coherent communication by mixing the input signal with a high enough intensity signal from a local oscillator[21]. On the other hand, non-coherent communication may or may not be shot noise limited. Depending on received power, in the photodetector the current fluctuations have both thermal and shot noise terms $\langle \Delta i^2 \rangle = (2q\langle i \rangle + kT/R)W$, $\langle i \rangle = qP/\hbar\omega$, and $R$ is the receiver impedance. For low enough optical signal power, thermal noise dominates the photo-current fluctuations and we get $SNR_e \sim P^2$, but as the optical power increases however shot noise dominates and we get $SNR_e = P/(2\hbar\omega W) \sim P$. These results are in agreement with both power[22] and SNR dependence[23] of the capacity of classical non-coherent optical communication in linear channels.

Similar results are shown in Fig. 2 for FCA inter-symbol interference (ISI) in single-channel transmission. The same waveguide dimension and carrier lifetime are used, only the symbol period is now 100 ps, so that approximately a total of 9 past symbols contribute to the ISI. Again, the linear regime of increasing capacity is interrupted, at a peak capacity of ~ 10.6 bits/s/Hz and now the capacity fall-off rate from the peak is twice the rate of increase in the linear regime. This steeper fall-of capacity, as compared to the WDM case, is due to the steeper increase of ISI with input intensity. A remarkable observation in both WDM and single-channel cases is that noise due to nonlinear loss (TPA and FCA) limits the capacity at power levels well below the level at which the nonlinear loss itself becomes significant (see Fig. 3).

In the case of non-coherent WDM, we assumed the current practice, in which channels are coded separately, and to the degree that each channel is subject to a random delay relative to the others, this approach is unavoidable. However, if the channels can be synchronized because the waveguide dispersion can be made low enough to guarantee an insignificant walk-off, joint coding is possible and it allows overriding the peaking of the capacity with power by sacrificing a fixed amount of bit rate (by use of a constant weight code).

## Figures

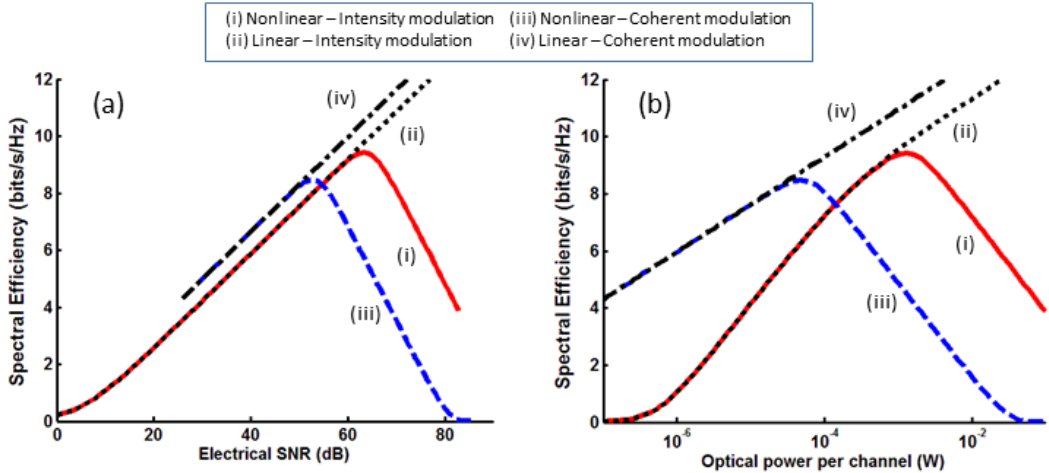

**Figure 1** (a) Channel capacity of silicon versus signal-to-noise ratio for wavelength division multiplexed (WDM) communication. Capacities for both intensity modulation (incoherent signaling) and coherent modulation are shown. For the coherent case, the capacity is for one of the two quadrature channels, in other words, the total capacity for both in-phase (I) and quadrature (Q) channels will be twice the value shown. The dotted lines show the capacity for a linear channel with additive white Gaussian noise (AWGN). (b) The same capacity versus optical power. In the case of WDM communication, the dominant source of fluctuations is the TPA induced cross talk between channels. 10 WDM channels (N = 10), W = 4 Gbps per channel, waveguide length $L_{wg}$ = 1 cm, effective area $A_{eff}$ = 0.5 µm$^2$, linear loss 0.5 dB/cm, electrical receiver impedance $R$ = 50 Ω.

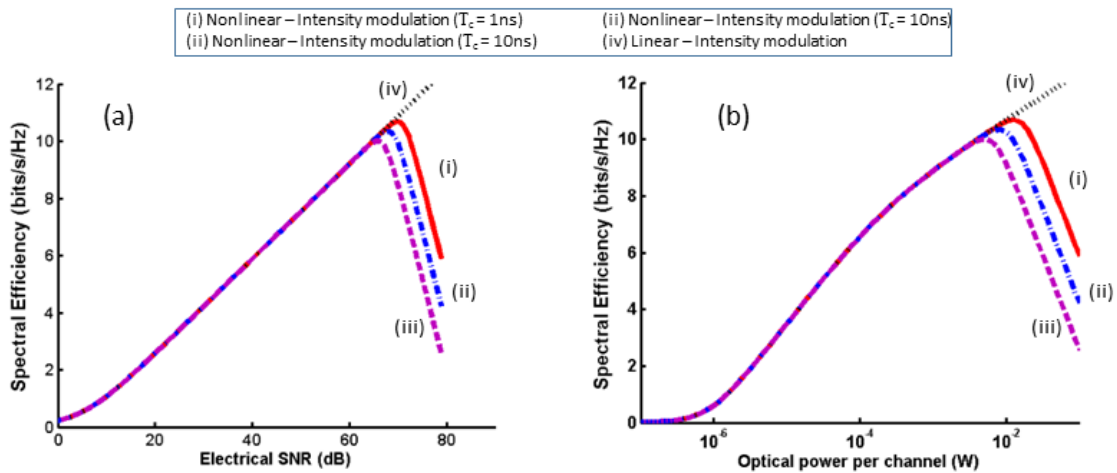

**Figure 2** (a) Channel capacity of silicon versus signal-to-noise ratio for single-channel communication. The dotted line shows the capacity for a linear channel with additive white Gaussian noise (AWGN). (b) The same capacity versus optical power. 10 Gbps data rate, 1ns lifetime, $L_{wg}$ = 1 cm, effective area $A_{\text{eff}}$ = 0.5 µm², linear loss 0.5 dB/cm, electrical receiver impedance $R$ = 50 Ω. The peak capacity has a $\frac{1}{10}log_2(\tau)$ dependence with respect to the carrier lifetime (see eqn. (11)).

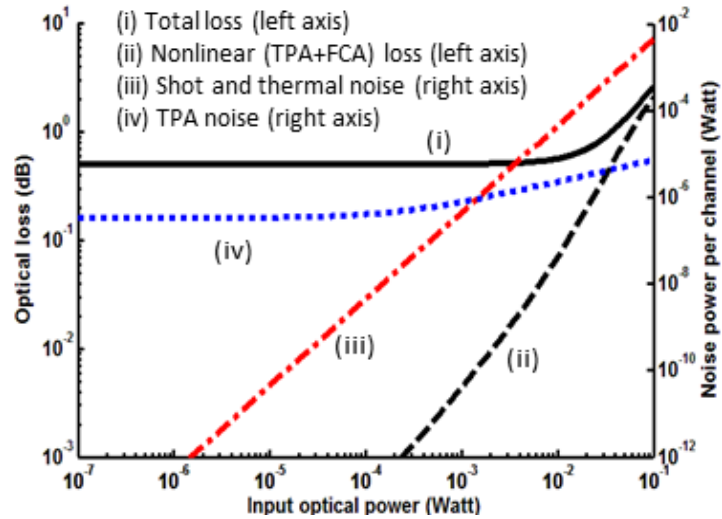

**Figure 3** Loss and fluctuation in a silicon waveguide channel with WDM communication. Two Photon Absorption (TPA) plus Free Carrier Absorption (FCA) loss, and total waveguide loss (including passive insertion loss) are shown (left y-axis). Also, shot and thermal noise powers, as well as nonlinear TPA noise are plotted (right y-axis). 1 ns lifetime, $L_{wg}$ = 1 cm, effective area $A_{\text{eff}}$ = 0.5 µm², linear loss 0.5 dB/cm, electrical receiver impedance $R$ = 50 Ω.

## Methods

### Channel parameters

This section provides the channel parameters that are used in the log-normal model (proofs will be supplied elsewhere). The mean total loss in the waveguide, due to linear losses, TPA and FCA, is $l$. The loss fluctuation (standard deviation) parameter is decomposed as $s^2 = s_{FCA}^2 + s_{TPA}^2$, where $s_{FCA}$ is ignored in WDM and $s_{TPA}$ is ignored in the single-channel calculation. For a single channel or multiple-channel WDM, $s_{TPA}$ and $s_{FCA}$ are respectively:

$$s_{FCA}^2 = \frac{P^4}{P_{FCA}^4}, \quad P_{FCA} = (P'_{TPA} P'_{FCA})^{1/2} \left(\frac{\tau}{10T}\right)^{1/4} \qquad (7)$$

$$s_{TPA}^2 \cong \frac{P^2}{P_{TPA}^2}, \quad P_{TPA} = (L_{eff}/L_{wg}) P'_{TPA}/N_C^{1/2} \qquad (8)$$

with $P$ the optical power per channel at the input of the waveguide, $P'_{TPA} = A_{eff}/(\beta L_{wg})$ and $P'_{FCA} = (A_{eff}/\sigma)(2\hbar\omega/\tau)$. The effective area of the waveguide is $A_{eff}$, and $L_{eff} = (1 - e^{-a_{LIN} L_{wg}})/a$ when $a_{LIN}$ is the linear loss coefficient.

### Capacity lower bound for single channel limited by FCA

By assuming that the power at the input of the waveguide is exponentially distributed, a very good lower bound is obtained. Directly in the lower bound expression (1), use the prescribed log-normal channel model to obtain (notations $\langle X \rangle = P$):

$$C_{LB}(\text{bits/s} \cdot \text{Hz}) = \frac{1}{2} \log_2 \left[ \frac{1 + \frac{e}{2\pi} \frac{e^{-2l} P^2}{P_N^2}\left(1 + 2\pi e^{-1-2\gamma} \frac{P^4}{P_{FCA}^4}\right)}{1 + \frac{2e^{-2l} P^2}{P_N^2} e^{(P/P_{FCA})^4}(e^{(P/P_{FCA})^4} - 1)} \right] \qquad (9)$$

This expression has units bits/(s·Hz), whereas expression (1) (after multiplied by $\log_2 e$) has units bits per channel use. To convert to our current normalization, we note that in non-coherent communication, an optical bandwidth $W$, results in an electrical bandwidth $W/2$, and therefore the rate of independent samples of the signal (channel uses) per second equals by Nyquist's theorem to $1/(2(W/2))$. The peak capacity is obtained approximately when $P^6 = e^{2l} P_{FCA}^4 P_N^2/4$. In the shot noise limited regime, $P_N^2 = P P_{shot}$, and the peak capacity occurs at:

$$P_{peak} = e^{2l/5} (P_{FCA}^4 P_{shot}/4)^{1/5} \qquad (10)$$

with a peak value of:

$$C_{peak} = \frac{1}{2}\log_2\left(\frac{e}{3\pi 4^{1/5}}\right) + \frac{2}{5}\log_2\left(\frac{e^{-2l}P_{FCA}}{P_{shot}}\right) \qquad (11)$$

### Capacity bound for coherent signaling

This bound was derived for WDM transmission in fiber[10] , but is equally applicable to silicon, since the responsible mechanism, cross-phase modulation (XPM), is also present in silicon.

Let us consider the transmission of $N_C$ channels, centered around a wavenumber $k_o = 2\pi/\lambda$, each of bandwidth $W$, in a fiber of length $L$ with a Kerr variation for the index of refraction $n = n_o + n_o I$, where $I$ is the input intensity. The per channel input power at which the nearby channels induce noise equal to the additive noise present is $P_{max} = (P_o^2 P_n/2)^{1/3}$. In this expression, $P_n = \hbar\omega W$ is the channel shot noise, and $P_o$ is the input power at which noise due to nonlinear effects appear, equal to $P_o = 1/\sqrt{2g^2 \ln(N_C/2) LL_{min}}$, where $L_{min} = \min(L, L_d)$ (here $L$ is the waveguide length). In these expressions $g = n_2/(k_o A_{eff})$, and $L_d$ is the dispersion length of the medium : $1/L_d = W^2 Dc/\lambda^2$, given in terms of the dispersion parameter $D$ (units ns/(km x nm)). Dispersion can be tailored for a wide variety of silicon waveguide parameters[24] , but we can consider a typical value $L_d$ ~ 1 cm (therefore $L_{min}$ ~ 1 cm). Considering that in silicon $n_2$ is 100 times higher than in fibre, and that effective areas of silicon waveguides are close to $0.5\ \mu m^2$ (200 times lower than fibre[25], gives $P_o$ = 2.8 W; and for 2 GHz channels, the shot noise power has a value of 258 pW, placing the onset of capacity limitation to 1 mW/channel.

### Four Wave Mixing

The refractive index $n$ depends on the intensity in the form of $n = n_o + n_2 I$. Here, $n_o$ is the linear index and $n_2$ is a constant characterizing the strength of the nonlinearity. The nonlinear coefficient $n_2$ in silicon is large rendering nonlinearity observable even in short distances on a chip. In the case of single channel (non WDM) the principle effect is self-phase modulation (SPM) where the refractive index and hence the phase depends on the intensity. In multi-channel (WDM) communication, this nonlinearity creates two other effects. The phase in a given channel is influenced by intensities in other channels in a process called cross-phase modulation (CPM). Also, electric field amplitudes in two channels influence the phase in the 3rd channel in a process called four-wave mixing (FWM). FWM depends on phase matching between the 3 fields and is curbed, naturally or intentionally, via dispersion. Since it can be avoided, FWM is not be considered here. FWM was also ignored in calculating the capacity of optical fibers[10].